\newif\ifdraft
\draftfalse

\documentclass[letterpaper,twocolumn,10pt]{article}
\usepackage{caption}
\usepackage{usenix,epsfig,endnotes}
\usepackage{tikz}
\usepackage{listings}
\usepackage{subcaption}

\usetikzlibrary{positioning,fit,arrows.meta,backgrounds,shapes.geometric}

\begin{document}

\date{}

\title{\Large \bf SafeLLVM: LLVM Without The ROP Gadgets!}

\author{
{\rm Federico Cassano}\\
Northeastern University
\and
{\rm Charles Bershatsky}\\
Northeastern University
\and
{\rm Jacob Ginesin}\\
Northeastern University
\and
{\rm Sasha Bashenko}\\
North Broward Preparatory School
}

\maketitle

\thispagestyle{empty}

\subsection*{Abstract}
\label{sec:abstract}


Memory safety is a cornerstone of secure and robust software systems, as it prevents a wide range of vulnerabilities and exploitation techniques.
Among these, we focus on Return-Oriented Programming (ROP). ROP works
as such: the attacker takes control of the program's execution flow via a memory 
corruption attack. Then, takes advantage of code snippets already in the
program's memory, dubbed "gadgets," to achieve the attacker's desired effect. 

In this paper, we introduce SafeLLVM, an approach to minimize the number of gadgets in x86-64 binaries compiled with the LLVM infrastructure \cite{llvm}.
Building upon the techniques outlined in previous works, we implement a series of passes within the LLVM compiler's backend to minimize the number of gadgets present, thus preventing ROP attacks.
We evaluated our approach by compiling a number of real-world applications, including cJSON, zlib, curl, and mimalloc.
The results from said evaluations demonstrated our solution's ability to prevent any form of ROP on the binaries compiled with SafeLLVM while maintaining the same functionality as the original binaries.

\section{Introduction}
\label{sec:intro}

Stack-based memory corruption vulnerabilities are a major security 
issue in modern software. Through such vulnerabilities, attackers can
gain control over the execution of a program, potentially leading to
the compromise of the entire system running it. Originally, these
vulnerabilities were exploited by overwriting the return address of a
function call with the address of a malicious payload inserted in the
stack region of the program's memory \cite{smashing-stack-profit}.
When the function returned, the execution would continue from the
malicious payload, allowing the attacker to execute arbitrary code
with the privileges of the compromised program. This kind of 
attack has been mitigated by the introduction of non-executable stack
protection mechanisms (NX on Linux or DEP or Windows) \cite{non-exec-stack},
which prevent the execution of code in the stack region. However,
these mechanisms are not effective against other types of memory
corruption vulnerabilities, such as those that utilize code
already present in the program's memory \cite{ret2libc}. Return-oriented programming
(ROP) is a technique that exploits such vulnerabilities by chaining
snippets of existing code in the program's memory to form a malicious
payload \cite{return-oriented-programming}. These snippets, called gadgets, are mostly
small pieces of code that perform a single operation and end with a
\textit{free-branch} instruction \cite{gfree}, such as \texttt{ret} or
\texttt{jmp \%reg}. With a sufficient number of gadgets, an attacker
can construct a malicious payload that performs a sequence of operations
that would be semantically equivalent to the execution of arbitrary
code, even reaching turing completeness \cite{rop-turing-complete}.
This brings us to a key idea: if we can remove the
gadgets from the program's memory, ROP would not be possible. 
G-Free is a technique that expands on this idea and removes usable
gadgets for x86 32-bit binaries \cite{gfree}.

In this paper, our goal is to modernize the G-Free technique by introducing a set of optimizations in the LLVM compiler's backend. These optimizations aim to reduce the presence of gadgets in program memory. Our primary focus during this process is set on x86 64-bit binaries, given that they are a contemporary standard in most operating systems. We also focus on performance and reliability, as we want to minimize
the impact of our technique on the original program's
functionality. Furthermore, we want to make our tool easily
portable to various operating systems and compilers. We achieve this
by implementing our passes in the LLVM infrastructure, which
is a widely used compiler framework that is portable to many
architectures, operating systems, and programming languages.

This paper provides the following contributions:
\begin{itemize}
  \item Exploration of the techniques used in G-Free,  tailored for x86 64-bit
  binaries, as well as newer techniques that are more effective
  in the 64-bit context.
  \item Implementation of several passes in the x86 LLVM backend 
  that utilize the techniques proposed in the previous point, and 
  the public release of our tool's, SafeLLVM, source code. 
  \endnote{The source code of SafeLLVM is available at \url{https://github.com/cassanof/safe-llvm}}.
  \item We assess our technique's effectiveness by comparing the number of gadgets in a program's memory before and after our passes are applied, as well as evaluating the tool's performance and reliability through testing it on various real-world test suites.
\end{itemize}

\section{Related Work}
\label{sec:related}

A lot of work has been done in the field of ROP mitigation and 
other memory corruption vulnerabilities. The techniques that prevent the exploitation of memory corruption vulnerabilities and the techniques that bypass said protections are two sides of the same coin. In this section we will discuss them both, highlighting the
techniques that are most relevant to our work.

\subsection{Stack Canary}
\label{sec:related:canary}

A popular technique to mitigate ROP attacks is the use of stack canaries:
random values that are placed on the stack before a function
call and are checked before the function returns. \cite{stackguard}.
If the value has been modified the program will halt, leaving the attacker with no
control over the execution of the program. However,
if the attacker possesses the canary value (e.g. from a memory
leak), they can bypass the protection. Furthermore, on 32-bit
binaries, the canary value is small enough such that it can be 
brute-forced in a reasonable amount of time \cite{preventing-canary-bruteforce}.
Additionally, more sophisticated attacks have been devised that
bypass the canary, without the need to know
its value \cite{canary-bypass}.

\subsection{Address Randomization}
\label{sec:related:randomization}

Address randomization is a technique that randomizes the memory
addresses of the program's code and data segments, making it
harder for an attacker to predict the addresses of the gadgets
they need to exploit a vulnerability. Sometimes, this requires the attacker to find a memory leak in the program \cite{pic-reuse-effectiveness-aslr}.

\subsubsection{Address Space Layout Randomization (ASLR)}

The simplest form of address randomization is ASLR, which 
shifts the base address of the program's memory by a random
amount at each execution \cite{aslr}. In the modern Linux kernel, ASLR is
enabled globally by default, randomizing the base address of segments
of all processes. However, if an attacker possesses the address of a
gadget, they can bypass ASLR by calculating the offset between the
gadget's address and the base address of the segment it belongs to,
and then adding the offset to the base address of the segment of the
attacker's choice \cite{aslr-bypass-leak}. 
Alternatively, ASLR can be easily
bypassed by utilizing \texttt{jmp \%rsp} gadgets, which
allow the attacker to jump to the stack region where another
gadget resides \cite{aslr-bypass-jop}. Finally, as 
ironic as it may sound, it has been shown that ASLR can aid 
exploitation of memory corruption vulnerabilities by introducing
wild-card ROP gadgets in PIC environments \cite{badaslr}.

\subsubsection{Position Independent Code (PIC)}

Position Independent Code (PIC) is a technique that allows the
program to be loaded at any address in memory, and still be able
to execute correctly. This is achieved by replacing absolute
addresses with relative addresses and using
indirect jumps and calls \cite{pic-disassembly}.
In Linux environments, Position Independent Executables (PIE)
can be compiled with the \texttt{-pie} flag, which enables
PIC for the program's code and data segments.
In contrast to ASLR, which randomizes the base address of the
program's memory, PIC randomizes the addresses of the program's
code and data segments. This makes it even harder for an attacker
to predict the addresses of the gadgets \cite{pie-makes-exploit-hard}. However, this still
doesn't make it impossible for an attacker to exploit gadgets
that are already present in the program's memory,
as the robustness of PIC is still limited to 
the source of randomness that ASLR uses \cite{aslr-robustness}.
For example, in Linux environments, PIE binaries get loaded as shared libraries and are all located side-by-side in memory, which allows an attacker to predict the addresses of the gadgets in the shared libraries by calculating the distance between them. \cite{offset2lib}.
Finally, PIE has been shown to drastically decrease the
performance of compiled binaries due to the overhead of
the extra instructions that are required to implement PIC
\cite{pie-performance}.

\subsection{Control Flow Integrity}
\label{sec:related:cfi}

Control Flow Integrity (CFI) is a technique that prevents the
tampering with the control flow of a program by checking the
integrity of a program's control flow graph at runtime \cite{cfi}, 
making it harder for an attacker to jump to a gadget
that is not explicitly allowed by the control flow graph.
Recently, Intel has introduced a new kind of CFI,
Control Flow Enforcement Technology (CET) \cite{cet-security-analysis}, which
is a hardware-based CFI that is supported by the latest
Intel processors. CET utilizes a shadow stack to keep track
of the return addresses of the functions that are currently
on the stack, and it checks the integrity of the stack at
each function call and return. The Linux kernel is slowly
moving towards integrating with CET, and it has been shown that CET can
mitigate ROP attacks, but not prevent them completely
\cite{survive-hacfi}.
Additionally, it has been shown that CFI can be bypassed by
leveraging speculative execution and write-what-where attacks
of control-flow data, where the information required for 
the write-what-where attack is leaked by the speculative execution
of a gadget acting as a side channel send \cite{mambretti2021bypassing}.

\subsection{G-Free}
\label{sec:related:gfree}

\begin{figure}[t]
  \centering
  \begin{tikzpicture}
    \node (b1) at (0,0) {41};
    \node (b2) at (1,0) {33};
    \node (b3) at (2,0) {57};
    \node (b4) at (3,0) {30};
    \node (b5) at (4,0) {c0};
    \node (b6) at (5,0) {c2};
    \node (b7) at (6,0) {05};
    \node (b8) at (7,0) {c3};
    \path[] (b1) -- (b2);
    \path[] (b2) -- (b3);
    \path[] (b3) -- (b4);
    \path[] (b4) -- (b5);
    \path[] (b5) -- (b6);
    \path[] (b6) -- (b7);
    \path[] (b7) -- (b8);
    \node[fit=(b3) (b4) (b5) (b6) (b7) (b8), label={[text=red, align=center]Unaligned Gadget\\
        \texttt{
          push \%rdi; xor \%al, \%al; ret \$0xc305
        }
      }, thick, draw=red, inner sep=3mm] (unaligned) {};
    \node[fit=(b1) (b2) (b3) (b4) (b5) (b6) (b7) (b8), label={[yshift=-2cm, text=blue, align=center]Aligned Gadget\\
      \texttt{
        xor 0x30(\%r15), \%edx; rol \$0x5, \%dl; ret
      }
    }, thick, draw=blue, inner sep=1mm] (aligned) {};
  \end{tikzpicture}
  \caption{Byte sequence encoding both a unaligned and an aligned gadget.}
  \label{fig:related:bseq}
\end{figure}

G-Free is a technique that attempts to remove all gadgets from
the program's memory, making it impossible for an attacker to
exploit a code reuse vulnerability \cite{gfree}. G-Free
works by replacing all the gadgets in the program's memory
with semantically equivalent code that does not contain 
usable free-branch instructions. To do this, G-Free 
makes a distinction between \textit{aligned} and \textit{unaligned} gadgets.
As shown in Figure \ref{fig:related:bseq}, aligned gadgets are gadgets that contain the same instructions
that would be executed if the gadget was executed normally,
while unaligned gadgets are gadgets that contain additional
instructions that are not normally executed, which are
fetched by the processor due to a misalignment of the 
program counter. Because aligned gadgets are part of the
program's normal execution flow, they cannot be removed
without changing the program's semantics. Instead, G-Free
protects them by enforcing the integrity of the program's
control flow around them.
Moreover, since unaligned gadgets are not normally executed, they can be
simply removed without changing the program's semantics. G-Free
manages to remove all the unaligned gadgets from the program's
by replacing the instructions with semantically equivalent
instructions that do not contain any free branches.
G-Free's main drawback is that it performs transformations
on the program's assembly code, and therefore it may not be
compatible with compilers of some programming languages.
Furthermore, G-Free was originally implemented for x86-32 binaries, and
it has not been extended to support x86-64 binaries. In this 
paper, we extend G-Free to support x86-64 binaries, doing so
by introducing new LLVM backend passes that perform
similar transformations to those of G-Free.
These passes can act upon the LLVM MIR, and therefore
can be used with any compiler that targets LLVM.

\section{Methodology}
\label{sec:methodology}

Following the methodology described by G-Free \cite{gfree}, we introduce two core 
groups of techniques that remove or render useless
all the gadgets from the
program's memory. The first group's goal is to protect
aligned gadgets, and the second group's goal is to remove
unaligned gadgets. 

\subsection{Protecting Aligned Gadgets}
\label{sec:methodology:aligned}

Aligned gadgets are part of the program's normal execution
flow, and therefore they cannot be removed without changing
the program's semantics. Instead, we must protect them by
enforcing the integrity of the program's control flow around
them. By protecting aligned gadgets, we ensure that the
instructions surrounding the gadgets are executed in the
correct order, and that the gadgets are executed only
in their valid context.

\subsubsection{Encrypting the Return Address}
\label{sec:methodology:aligned:encrypt}

\begin{figure}[htpb]
  \centering
  \begin{lstlisting}[frame=tb, captionpos=b, label={lst:methodology:aligned:encrypt}]
0000000000001000 <add>:
    ;; encrypt return address
    1000: mov    %fs:0x28, %r11
    1009: xor    %r11, (%rsp)
    ;; start of the function
    100d: push   %rbp
    100e: mov    %rsp, %rbp
    1011: mov    %edi, -0x4(%rbp)
    1014: mov    %esi, -0x8(%rbp)
    1017: mov    -0x4(%rbp), %eax
    101a: add    -0x8(%rbp), %eax
    101d: pop    %rbp
    ;; decrypt return address
    101e: mov    %fs:0x28, %r11
    1027: xor    %r11, (%rsp)
    ;; return to caller
    102b: ret
  \end{lstlisting}
  \caption{Example assembly code of a function that uses the encrypt and decrypt subroutines.}
  \label{fig:methodology:aligned:encrypt}
\end{figure}

Similar to how a stack canary acts as a sentinel
value on the stack, we can encrypt the return address 
of a function stored on the stack every 
time the function is entered, and then decrypt
it before returning from the function.
If the return address is not properly decrypted, then
the function will return to an unexpected address, and
the program will most likely crash. This
ensures that every single function containing a
\texttt{ret} instruction will need to be executed in
full, which prevents attackers from exploiting
code reuse vulnerabilities by jumping to the middle
of a function.

In our implementation we utilize the stack canary secret stored at \texttt{fs:0x28} as
as source of randomness to encrypt the return address, but
one may use any other source of randomness. We
chose to utilize the stack canary secret as it is already
present in memory, and it only requires a single read
instruction to retrieve it. Meanwhile, G-Free utilizes a different
source of randomness, and it requires a call to a
subroutine reading a sequence of bytes from the \texttt{/dev/random}
device \cite{gfree}. This makes the implementation of G-Free more complex,
and relies on the device being available
on the system, which is typically only available on Unix-like
operating systems.

As shown in Figure \ref{fig:methodology:aligned:encrypt}, 
the encryption and decryption subroutines utilize the 
same sequence of two instructions to encrypt and decrypt
the return address. The first instruction is a \texttt{mov}
instruction that reads the stack canary secret from memory
and stores it in a temporary register. The second instruction
is a \texttt{xor} instruction that performs an exclusive-or
operation between the temporary register and the return address
stored on the stack. The XOR operation is performed in-place,
and therefore it does not require any additional memory accesses.

\subsection{Removing Unaligned Gadgets}
\label{sec:methodology:unaligned}

Having dealt with aligned gadgets, we now turn our attention to unaligned gadgets. If an attacker were to
move the instruction pointer to the middle of an instruction, 
the processor would fetch and execute a sequence of unintended
instructions. An attacker could craft a program that would be
able to determine the exact sequence of instructions that would
be executed, and therefore the attacker could use this to
find gadgets that wouldn't normally be executed. For example,
the instruction \texttt{mov \$0xc3, \%r11}, encoded as
\texttt{49 c7 c3 c3 00 00 00}, would normally be executed
as a single instruction. However, this instruction hides
two \texttt{ret} instructions, encoded as \texttt{c3}, which
would be executed if the instruction pointer were to be
moved to the middle of the instruction.

Our primary solution for aligned gadgets is to add protection 
to the gadget that would act at runtime; this would protect the code without modifying any of the existing instructions.
However, for unaligned gadgets, we can \textit{statically} remove the gadgets
from the program's memory by modifying the program's instructions directly
and substituting them with semantically equivalent instructions that do not
contain any free branches.

Critically, we need to remove \textit{all} ret-based free-branch instructions in unaligned gadgets, 
which includes the following instructions:
 
\begin{itemize}
  \item \texttt{ret} (0xc3)
	\item \texttt{ret imm16} (0xc2)
	\item \texttt{retf} (0xcb)
	\item \texttt{retf imm16} (0xca)
	\item \texttt{iret} (0xcf)
\end{itemize}

Because only one byte is needed to encode the actual free branch instruction, a free branch could be hiding anywhere throughout a larger instruction - in the instruction sequence, the MOD R/M, SIB bytes, or any arguments (displacements, immediates, or registers). For example, the following byte sequences all encode free-branch instructions:

\begin{itemize}
	\item \texttt{\textbf{c3}}
	\item \texttt{13 37 \textbf{c3}}
	\item \texttt{13 \textbf{c3} 37}
\end{itemize}

This means that in removing all unaligned gadgets, 
we need to ensure that we check every byte of the target instruction
for the presence of a free branch.

\subsubsection{Restoring Alignment With NO-OP Instructions}
\label{sec:methodology:unaligned:nops}

Having added an XOR protection to every aligned gadget, we now find ourselves wondering how we can defend against any unaligned gadgets that may be present in our modified function. Changing the code would, again, run the risk of affecting its functionality. As such, we prepend each aligned \texttt{ret} with a \texttt{nop} sled. 

A \texttt{nop} sled is a series of no-op instructions, each of which is only a single byte.
When the processor fetches and executes a no-op instruction, it will simply
increment the instruction pointer by one byte, and continue to the next instruction.
While \texttt{nop} sleds are often found in the context of exploitation
\cite{momotcreatively, kolesnikov2005advanced, stevens2011malicious},
the sole purpose of this application is to ensure that by the end
of the sled, the instruction pointer is correctly aligned.

We follow the same methodology as G-Free in determining the length of the sled, where 
the length of the sled is calculated as the maximum length of an instruction. Additionally, we add one byte due to the possibility of a \texttt{nop} instruction being read as a MOD R/M or SIB byte instead.
In the original implementation of G-Free, a \texttt{nop} sled of length 9 was used since
the maximum length of a 32-bit instruction is a mere 8 bytes.
We came up with a sled length of 16 bytes, which is the maximum length of an instruction on a x86-64 processor,
plus one byte to accommodate for the possibility of a SIB or MOD R/M byte.

A common concern in using sleds like this for alignment is the possible performance costs.
As such, a seemingly intuitive addition would be to add a \texttt{jmp} instruction at the start of
the \texttt{nop} sled, which would allow execution with a properly aligned instruction pointer to
skip over the 16 instructions that are prepended to every \texttt{ret}. 
However, since modern processors are superscalar, the processor is able to execute multiple
instructions in parallel \cite{shen2013modern}, so the performance cost of the sled is negligible. Therefore, we
opt to not include the \texttt{jmp} optimization, as it would add an additional instruction to every
aligned gadget, which would increase the size of the code and therefore the size of the code cache \cite{hu2006approach}.

\subsubsection{Re-encoding Immediate Values}
\label{sec:methodology:unaligned:reencode}

\begin{figure}[htpb]
\centering
\begin{subfigure}{.2\textwidth}
    \begin{minipage}[t]{.5\linewidth}
      \begin{lstlisting}
mov $0xc3, %rax
      \end{lstlisting}
  \end{minipage}
  \caption{Instruction before the transformation. 0xc3 is the opcode for \texttt{ret}, and
    it is being moved into the \texttt{\%rax} register.}
  \label{fig:methodology:unaligned:reencode:before}
\end{subfigure}
\hfill
\begin{subfigure}{.2\textwidth}
    \begin{minipage}[t]{.5\linewidth}
        \begin{lstlisting}
mov $0x62, %r11
add $0x61, %r11
mov %r11, %rax
        \end{lstlisting}
    \end{minipage}
    \caption{Sequence of instructions after the transformation. 0xc3 is divided into 0x62 and 0x61.}
  \label{fig:methodology:unaligned:reencode:after}
\end{subfigure}
\caption{Example of dividing an immediate value into two smaller values, and
reconstructing the original value at runtime.}
\label{fig:methodology:unaligned:reencode}
\end{figure}

Instructions that contain immediate values may encode one 
or more gadgets. For example, the \texttt{mov} instruction
in Figure \ref{fig:methodology:unaligned:reencode:before} contains
the \texttt{ret} gadget in its immediate value, as the instruction
is assembled as \texttt{48 c7 c0 \textcolor{red}{c3} 00 00 00}, where \texttt{c3} is the
opcode for \texttt{ret}. To eliminate this gadget, we can divide the 
immediate into two parts and reconstruct the original value at runtime
by adding the two parts together using a temporary register (see Figure \ref{fig:methodology:unaligned:reencode:after}).
The original \texttt{mov} instruction that moves the immediate is replaced
with one that moves the temporary register into the destination register.

When we divide the immediate into two parts, we need to make sure
that neither of the two parts can encode a free-branch gadget.
For example, the \texttt{2a \textcolor{red}{c3} 85} sequence contains the \texttt{c3} byte,
which is the opcode for \texttt{ret}. If we naively divide the immediate 
by two, we will be left with \texttt{15 61 \textcolor{red}{c2}} and \texttt{15 61 \textcolor{red}{c3}}, which
both contain a free-branch (\texttt{c2} and \texttt{c3} respectively).
A solution to this problem is to use an algorithm that divides the immediate
into two parts such that the sum of the two parts is equal to the original
immediate value and the two parts do not contain any free-branch gadgets.
A robust algorithm for this problem can be described utilizing satisfiability
modulo theories \cite{smt}.

G-Free utilizes a very similar technique to ours, where the immediate 
value is reconstructed at runtime \cite{gfree}. However, G-Free does not
generalize the process and utilizes a different sequence of instructions for each
instruction. In contrast, we have implemented a generic solution that
can be applied to any instruction that contains an immediate value.

\subsection{LLVM Backend Passes}
\label{sec:methodology:llvm}

To implement the techniques described in Section \ref{sec:methodology},
we introduce multiple LLVM backend passes that act upon the LLVM MIR \cite{llvm-mir}.
For implementing the passes, we used LLVM 16 as a base and modified the code of the x86 target machine to call our 
passes at the appropriate time. Currently, we have implemented
two core passes.

\subsubsection{\texttt{SafeReturnMachinePass}}
\label{sec:methodology:llvm:safe-return}

\texttt{SafeReturnMachinePass} is responsible for
implementing the encrypt/decrypt subroutines described in
Section \ref{sec:methodology:aligned:encrypt}. The pass is called
before emitting the machine code for a function (\texttt{X86PassConfig::addPreEmitPass}), as we want to
transform the MIR after the LLVM optimizations have been applied
and the prologue and epilogue for every function have been generated.
The pass inserts the encryption subroutine at the beginning
of every function that contains a \texttt{ret}, and
the decryption subroutine before every \texttt{ret} along with a sled of \texttt{nop} instructions, as described in Section \ref{sec:methodology:unaligned:nops}.

\subsubsection{\texttt{ImmediateReencodingMachinePass}}
\label{sec:methodology:llvm:immediate-reencoding}

\texttt{ImmediateReencodingMachinePass} is the pass that performs the re-encoding of immediate values that encode a free-branch instruction,
as described in Section \ref{sec:methodology:unaligned:reencode}. The pass is called
before register allocation (\texttt{X86PassConfig::addPreRegAlloc}), as we will need to
allocate virtual registers to store the re-encoded immediate values, such
that registers will not be clobbered by the re-encoding process.
The pass scans the MIR for instructions that contain a free-branch-encoding immediate value as 
an operand and it re-encodes the immediate value dynamically as described in Section \ref{sec:methodology:unaligned:reencode}.
We only re-encode immediate values of instructions in the \texttt{ri}, \texttt{ri8}, and \texttt{ri32} classes,
as these are the only classes that contain instructions that can be used to encode a free-branch instruction in their immediate operand.
In order to re-encode the \texttt{ri*} instructions, we will need to convert them to their \texttt{rr} counterparts,
as the immediate value will be replaced with a virtual register. Currently, we only re-encode 
these following instructions: \texttt{add}, \texttt{sub}, \texttt{and}, \texttt{or}, \texttt{xor},
\texttt{cmp}, \texttt{test}, and \texttt{mov}.

Our process to re-encode the immediate value utilizes the \texttt{add} instruction to
combine both parts of the immediate value into a single immediate value stored in a virtual register.
Because of this, our \texttt{add} instruction may modify the flags register, which may affect the
execution of the following instructions. Therefore, we insert a \texttt{pushf} instruction before
the \texttt{add} instruction, and we insert a \texttt{popf} instruction after the \texttt{add} instruction,
such that the flags register will be saved and restored before and after the \texttt{add} instruction.

\section{Results}
\label{sec:results}

In this section, we present the results of our evaluation of SafeLLVM, focusing on two key aspects: the reduction of ROP gadgets in then compiled binaries, and the performance impact of SafeLLVM on the respective test suites. Our goal is to demonstrate the effectiveness of SafeLLVM in mitigating ROP attacks while assessing any potential trade-offs between enhanced security and performance. We compiled 8 widely-used software projects --- zlib, cJSON, mimalloc, curl, surf browser, Suckless Terminal (ST), Doom Chocolate, and LittleFS --- Using SafeLLVM and analyzed the results in terms of these two aspects.  

The following subsections detail our findings, discussing the reduction in ROP gadgets and the performance of the compiled binaries when running their respective test suites.



\subsection{Reducing ROP Gadgets}%
\label{sub:Remaining ROP Gadgets}

We compiled each project with both LLVM (version 16) and SafeLLVM and compared the number of ROP gadgets present in their respective binaries. We also employ Ropper, a tool that automatically discovers ROP chains to determine if compilation through SafeLLVM can prevent the automatic discovery of ROP attacks. 

\begin{table}[h]
\centering
\begin{tabular}{l|cc|cc}
\hline
Toolchain & \multicolumn{2}{c|}{LLVM} & \multicolumn{2}{c}{SafeLLVM} \\
\cline{2-5}
& Gadgets & ROP Chain & Gadgets & ROP Chain \\
\hline
zlib & 1169 & yes & 194 & no \\
cJSON & 525 & no & 64 & no \\
mimalloc & 2014 & yes & 377 & no \\
curl & 1268 & yes & 166 & no \\
SURF & 343 & no & 105 & no \\
ST & 999 & no & 306 & no \\
Doom & 7735 & yes & 1528 & no \\
LittleFS & 414 & no & 60 & no \\
\hline
\end{tabular}
\caption{Comparison of ROP gadgets and ROP chain discoverability between regular LLVM and SafeLLVM compilation.}
\label{tab:rop_comparison}
\end{table}

As shown in Table 1, compiling with SafeLLVM as opposed to LLVM results in a significant reduction in the number of ROP gadgets across all tested projects. Moreover, when using Ropper, we observed that an exploitable ROP chain could not be found in any of the projects compiled with SafeLLVM, thus effectively increasing the security of the compiled binaries.  

\subsection{Compiled Binary Performance}%
\label{sub:Compiled Binary Performance}

To analyze the performance impact of SafeLLVM on the compiled projects, we ran their respective test suites and measured the average completion time across 10 runs for both regular compilation and compilation using SafeLLVM. This allows us to gauge the trade-off between the enhanced security provided by SafeLLVM and the performance of the resulting binaries. 

\begin{table}[h]
\centering
\begin{tabular}{l|cc|cc}
\hline
Toolchain & \multicolumn{2}{c|}{LLVM} & \multicolumn{2}{c}{SafeLLVM} \\
\cline{2-5}
& Tests & Time (ms) & Tests & Time (ms) \\
\hline
cJSON & 19/19 & 40 & 19/19 & 40 \\
mimalloc & 3/3 & 4,706 & 3/3 & 1,395 \\
LittleFS & 817/817 & 6,420 & 817/817 & 6,505 \\
\hline
\end{tabular}
\caption{Performance comparison of test suite completion time and test results between regular LLVM and SafeLLVM compilation.}
\label{tab:performance_comparison}
\end{table}

As shown in Table 2, SafeLLVM has a minimal impact on the execution times of the test suits for cJSON and LittleFS. Interestingly, the mimalloc project exhibits a significant improvement in the test suite completion time when compiled with SafeLLVM, reducing the average time from 4,706ms to 1,395ms. At this time we are unable to pin this performance improvement to a specific characteristic of SafeLLVM, and we plan to investigate this further in the future. 

Lastly, we ran the \textit{compiler-benchmark} suite on both C and C++ benchmarks with a function 
count of 30 and a depth of 4.
On this suite, we found that SafeLLVM has a negligible impact on the performance of the compiled binaries, with an average slowdown of 0.2\% for C benchmarks and 0.1\% for C++ benchmarks.

Thus, SafeLLVM effectively reduces the number of ROP gadgets and prevents the automatic discovery of ROP chains without causing substantial performance overhead or negatively affecting the functionality of the compiled projects. 

\section{Discussion}
\label{sec:discussion}

We have demonstrated that SafeLLVM is able to reduce the number of gadgets in a binary, and in most
cases, prevent the automatic generation of ROP chains. However, there are some limitations to our
approach and some potential avenues for future work, which we discuss in this section.

\subsection{Limitations}
\label{sec:discussion:limitations}

\subsubsection{Code that depends on the return address}
\label{sec:discussion:limitations:retaddr}
In our LLVM pass, we haven't adopted a mechanism
to detect the presence of any code that depends on accessing the 
return address. For example,
\texttt{setjmp} and \texttt{longjmp} are functions that save and
restore the return address of a function call
in order to implement non-local jumps, which unwind the stack
and return to a previous function call \cite{setjmp-longjmp}.
When using our technique, the return address accessed by \texttt{setjmp} is encrypted, and therefore
the \texttt{longjmp} call will not be able to restore the correct return address,
potentially leading to a crash.

\subsubsection{Stack Canary Leaks}
\label{sec:discussion:limitations:canary}
Our technique utilizes the stack canary secret (stored at \texttt{\%fs:0x28})
to encrypt the return address. This means that if the attacker can
leak the canary secret, they can also decrypt the return address, 
rendering our technique useless. A possible solution to this problem
is to create a unique key for every function call and store it in 
the stack frame of the function. This way, even if the attacker 
possesses a memory read primitive, they will not be able to decrypt
the return address since calling the same function twice will result
in two different keys. However, this method decreases the performance
of programs that make a large number of function calls dramatically.
The most efficient solution would be one that utilizes the \texttt{RDRAND \%reg}
instruction, which is a hardware-based cryptographically secure pseudo-random number generator \cite{intel-drng}.
Depending on the processor vendor, this instruction can take more than 1000 CPU cycles to execute \cite{fog2011instruction}, which is much
slower than the execution of a typical function call. However, such
a performance penalty may be acceptable when the goal is to protect
critical services that are not performance-critical.

\subsubsection{Inline Assembly}
Finally, our technique does not support inline assembly. This is because
the LLVM Backend does not provide constructs to traverse
the inline assembly code without implementing a custom assembler and 
assembly printer.
Inline assembly code containing unaligned gadgets would remain
untouched by our technique, and therefore would still be exploitable.
Moreover, if the inline assembly code contains a \texttt{ret} instruction,
our technique may encrypt the return address without decrypting it,
potentially leading to a crash.

\subsection{Future Work}
\label{sec:discussion:future}

To address the limitations listed in \ref{sec:discussion:limitations}, we plan to
implement a static analysis pass that marks machine functions that
depend on the return address. For each marked function, we will decrypt the return address
right before the basic block that utilizes it, and re-encrypt it right after.
Then, we will implement an optional pass that generates a unique key for every function call as 
described in \ref{sec:discussion:limitations:canary}. This pass will be optional since it 
will drastically decrease the performance of some programs.

Furthermore, our technique tries to minimize gadgets ending with the 
\texttt{ret} free-branch instruction, mitigating ROP. However, there are other
gadgets that can be used to perform other attacks, such as Jump-Oriented Programming (JOP),
which utilizes gadgets ending with \texttt{call \%reg} and \texttt{jmp \%reg}
free-branch instructions. In the future, we plan to extend our technique
to also remove these gadgets.

Implementing our own assembler would be a very complex task, but
it would allow us to support inline assembly and perform 
more complex analysis on the low-level IR that is generated by the compiler.
Therefore, we decided to leave this as future work.

Since the ARM architecture is starting to be used more and more,
and is heavily used in mobile and embedded devices, we plan to extend
our technique to the ARM architecture too. Additionally, we plan
to generalize our technique in order to be able to make it 
easy to implement for an arbitrary architecture.

{\footnotesize \bibliographystyle{acm}
\bibliography{main}}

\theendnotes

\end{document}